# Can Second-Order Nonlinearity in Metal-Dielectric Metamaterials Pave a Way Toward Elusive Photonic Time Crystals?

Jacob B Khurgin[1*]
[1]Department of ECE, Johns Hopkins University, Baltimore MD 21218 USA
[*]Corresponding author: jakek@jhu.edu

**Abstract**

While a tremendous amount of theoretical work has been dedicated to time-varying photonics, practical implementation in the optical range has relied almost exclusively on transparent conductive oxides , which remain severely constrained by being slow and highly lossy. To bypass these limitations, I propose an alternative platform utilizing ultrafast second-order nonlinearities within an epsilon-near-zero semiconductor-silver metamaterial. Due to the relatively low loss in silver and polarization-selective pumping, ultra-low pump absorption prevents thermal degradation, while absorption of the probe is sufficiently low to simultaneously permit the multi-cycle interaction lengths necessary for signal detection. I find that with pump powers of 100s of GW/cm$^2$ one can open a wide momentum bandgap of tens of percents and achieve net parametric gain in time, offering a robust pathway to realizing photonic time crystals in optical range.

Recent years have seen a surge of interest in time-varying phenomena in photonics that explore space-time duality—hence a number of temporal analogs to well-known spatial photonic structures (reflection, refraction, diffraction, etc.) have been translated into the temporal domain[1-4]. Let us not bother the reader with well-beaten, ubiquitous statements about how these phenomena will ostensibly revolutionize the world as we know it—from communications to data centers to AI to quantum computing—but simply accept as fact that if a great many important people in the photonics community deem time-varying phenomena important, they must indeed be important, if only for fundamental reasons. A quick disclaimer is in order: since this is a short letter rather than a review, I have not attempted to reference the entire vast trove of works in the field—pioneering, valuable and all-encompassing though they all may be.

All time-varying phenomena hinge on the possibility of strongly modulating a medium's optical properties—namely, its permittivity $\varepsilon$ and refractive index $n$. The effect most critically dependent on a large modulation depth is the photonic time crystal (PTC)[5-9], where modulating the dielectric permittivity at frequency 2ω opens a gap in momentum space $\Delta k \approx \omega\Delta\varepsilon / 2c\varepsilon$, as shown in Fig.1a accompanied by the signature exponential temporal growth of a probe and time reflected fields near frequency $\omega$.

$$
\begin{aligned}
E_{probe}(t) &\sim \exp\left[\frac{\omega}{2\varepsilon}\left(\frac{\Delta\varepsilon}{2}-\varepsilon^{"}\right)t\right]E_{probe,0}\cos(i\omega t-ikz) \\
E_{t-ref}(t) &\sim \exp\left[\frac{\omega}{2\varepsilon}\left(\frac{\Delta\varepsilon}{2}-\varepsilon^{"}\right)t\right]E_{ref,0}\cos(-i\omega t-ikz)
\end{aligned} \qquad (1).
$$

To observe net parametric gain, the permittivity modulation depth must overcome the inherent absorption loss dictated by the imaginary part of the permittivity $\varepsilon^{"}$. It is precisely this stringent requirement—achieving a large permittivity modulation at single-optical-cycle speeds while maintaining low optical loss—that has rendered the experimental realization of optical-frequency PTCs (and related time-varying phenomena) so exceptionally difficult to achieve in practice, despite no shortage of trying.

The difficulty of achieving a permittivity modulation on the order of $\Delta\varepsilon/\varepsilon \sim 0.1-1$ in transparent materials is both obvious and fundamental[10]. Given that virtually all known materials transparent in the visible and near-infrared regimes possess permittivities within a single order of magnitude of one another[11, 12], a permittivity change of this magnitude would effectively require reconfiguring the medium into an entirely new material—a process demanding an energy density commensurate with the cohesive energy density of the crystal lattice itself[13]. And since the modulation must occur at single-cycle speed, the only potentially feasible route to achieve it is optical (rather than the abstract index modulation by the "will power of the authors" so prevalent in the literature. Then, as supported by a vast body of experimental data on optical nonlinearities, the power densities required to drive such a large permittivity modulation become unsustainably high—reaching many tens of TW/cm$^2$, even without taking into account the inevitable saturation of the nonlinearity[14, 15].

And yet, researchers have discovered a way to circumvent this apparently impregnable obstacle and mitigate the stringent requirements by recognizing that it is the *fractional* rather than the *absolute* change in permittivity that determines the width of the momentum gap, as well as the strength of temporal reflection and refraction. Thus, when the baseline permittivity is small—as in so-called Epsilon-Near-Zero (ENZ) materials[16-19]—even a modest absolute permittivity change $\Delta\varepsilon$ can engender a relatively large fractional variation $\Delta\varepsilon/\varepsilon$.

This class of ENZ materials comprises transparent conductive oxides (TCOs) like ITO (indium tin oxide), AZO (aluminum zinc oxide), and others, whose dielectric constant can be expressed as:

$$\varepsilon(\omega) = 1 + \chi_b + \chi_f(\omega) \qquad (2)$$

where $\chi_b$ is the relatively dispersionless susceptibility of bound electrons (interband contribution), and

$$\chi_f(\omega) = -\omega_p^2 / \omega(\omega + i\gamma) \qquad (3)$$

is susceptibility of free carriers in the conduction band is the susceptibility of free carriers in the conduction band, where $\gamma$ is their momentum relaxation rate, and $\omega_p = \sqrt{Ne^2 / \varepsilon_0 m_c}$ is the plasma frequency determined by carrier density $N$ and effective mass $m_c$

When the operating frequency approaches the point where $1 + \chi_b + \mathrm{Re}\left[\chi_f(\omega)\right] \sim 0$ the real part of the permittivity approaches zero. By modulating the free-carrier parameters—either by exciting extra carriers into the conduction band to alter $N$, or by heating carriers within a nonparabolic band to alter $m_c^*$—one can obtain a substantial relative modulation of the net permittivity.

Now, the interband mechanism of altering the conduction band population is inherently slow—measured in hundreds of picoseconds—while intraband nonlinearity comprises two distinct components: a relatively small, ultrafast (instantaneous) contribution arising from electron oscillation in a nonparabolic band, and a substantially larger, slower contribution due to non-equilibrium carriers occupying higher-energy states in the band[20]. The latter is governed by the intraband relaxation (cooling) time of tens to hundreds of femtoseconds, which remains considerably longer than a single optical cycle[15, 21, 22].

While traces of the fast nonparabolic nonlinearity have been observed[23, 24], being a third-order process, its magnitude is strictly proportional to the instantaneous intensity of the pump pulse. The response can therefore be only as short as the optical envelope itself, meaning it cannot provide the sub-cycle, harmonic phase oscillations at $2\omega$ required to open a PTC momentum gap. Conversely, the slow thermal response is strong precisely *because* it is slow accumulating the overall pulse energy rather than responding to instantaneous optical power.

All of this is compounded by the heavy Ohmic loss inevitably present in ENZ media, where the "giant" nonlinearity itself relies directly on strong optical absorption[22]. Hence, while immense effort has been devoted to ENZ-based time-varying optics, the practical results thus far have been rather meagre. This is by no means intended to disparage the scientists involved in ENZ research—having been one of them myself, I should be the last person to throw stones—but at this juncture, I simply do not see a realistic path toward observing true photonic time crystals using ENZ materials as we currently know them.

Let us now summarize the trade-offs inherent to ENZ materials. The clear advantage is operating in the ENZ regime itself, which enables large relative modulation by balancing positive and negative contributions to the permittivity. The two major drawbacks, however, are the severe optical loss that causes lattice overheating, and the reliance on third-order nonlinearity ($\chi^{(3)}$)—which is perpetually trapped between being either ultrafast but far too weak, or strong but far too slow.

Furthermore, it is worth noting that in traditional ENZ media, both the reduction of permittivity toward zero and the optical nonlinearity itself are provided exclusively by free carriers. Bound carriers play no active role whatsoever, serving merely as the electronic glue that binds the ions together.

One potential step forward would be to harness the second-order nonlinearity ($\chi^{(2)}$) of the lattice instead of $\chi^{(3)}$, given that photonic time crystals are inherently parametric processes—and virtually all practical parametric devices operate using $\chi^{(2)}$ rather than $\chi^{(3)}$, despite the phase-matching challenges involved. However, ITO is centrosymmetric ($\chi^{(2)} = 0$), and AZO, like most TCOs, yields a negligible macroscopic $\chi^{(2)}$ due to its polycrystalline nature. On the other hand, epitaxially grown III-V semiconductors like GaAs, which possess massive non-resonant $\chi^{(2)}$ susceptibilities, cannot be doped heavily enough to reach the ENZ regime in the visible or near-infrared spectrum without destroying the crystal lattice. Furthermore, even if such doping were possible, the accompanying Ohmic loss from the high carrier concentration would lead to catastrophic overheating via pump absorption. Crucially, observing a true photonic time crystal requires maintaining parametric driving over many optical cycles, making this thermal buildup the primary experimental bottleneck.

Heavily doped materials, however, are not the only route to an effective ENZ medium. The same near-zero condition can be engineered by combining materials with positive and negative permittivities within a metamaterial[25, 26], where bound and free carriers remain spatially separated—for example, by embedding subwavelength metallic inclusions into a dielectric matrix. Because intrinsic momentum relaxation rates in metals can be lower than those in heavily doped TCOs, this spatial separation inherently reduces Ohmic losses.

Even greater performance gains can be achieved if the metamaterial is strongly anisotropic because while the ENZ regime is essential for the probe wave, it is entirely unnecessary (and indeed undesirable) for the pump. This ideal decoupling can then be realized in a transversely pumped geometry[27-29], where the pump wave propagates orthogonally to the probe. By spatially and functionally decoupling the free and bound carriers—letting the former shift the linear response into the ENZ regime while the latter provide the ultrafast permittivity modulation—each component can be optimized independently, potentially bringing us closer to the practical realization of photonic time crystals.

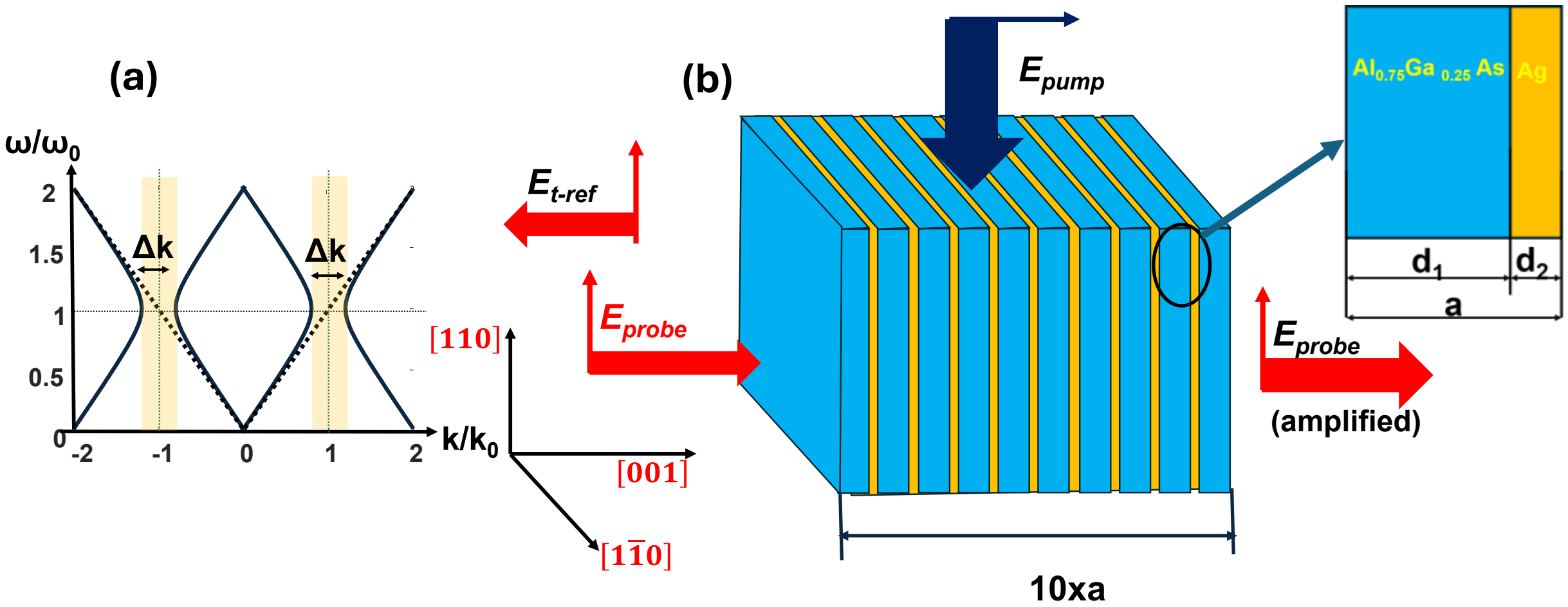

**Fig. 1.** (a) Sketch of the momentum-space bandgap in a photonic time crystal (PTC), with frequency and momentum plotted relative to their values at the center of the gap. (b) Geometry of the proposed WNZ metamaterial structure featuring large $\chi^{(2)}$ nonlinearity designed to yield a wide momentum bandgap

That being said, let us look at the possible implementations as shown in Fig.1b. To prevent inevitable questions of whether one can reliably fabricate such an arrangement, let us say right away that the goal of this work is to outline a pathway to PTC and provide order-of-magnitude estimates in a structure that can in principle be fabricated sooner or later. While a second inevitable question is naturally "what about phase matching?", the answer lies in employing a transverse pumping geometry with counterpropagating probe (signal) and time-reflected (idler) waves, where phase matching is automatic, as demonstrated in [27, 30] With that in mind, consider the structure depicted in Fig. 1. The structure consists of 10 periods with a width of $a = 200nm$. Each period comprises a thick layer $d_1$ of (001)-oriented $Al_{0.75}Ga_{0.25}As$ (direct bandgap energy of 2.35eV and a thin layer $d_2$ of silver. The choice of material composition is based on the necessity of avoiding two-photon absorption of the pump radiation, which for this example is chosen to be $\lambda_{pump} = 1150nm$(photon energy of 1.08eV). The strong pump radiation is incident upon the side facet, and, crucially, is polarized normal to the (001) plane in order to reduce its absorption in Ag. The facet itself is oriented in the natural cleavage plane (110).

The probe radiation then should have a wavelength at or near $\lambda_{probe} = 2\lambda_{pump} = 2300nm$ and be polarized either vertically or horizontally (along either [110] or [1$\bar{1}$0] directions). With this arrangement one can take advantage of the large tensor element of second order susceptibility of zinc blende lattice $\chi_{14}^{(2)} \equiv \chi_{xyz}^{(2)} \sim 100 pm/V$ the nonlinear polarization at probe frequency $\omega$ can be written as

$$P_{nl}(\omega) = 2\varepsilon_0 \chi_{14}^{(2)} E_{probe}^* e^{i\omega t - ikz} E_{pump} e^{-2i\omega t} + c.c.. \quad (4)$$

This polarization can generate time reflected counterpropagating idler wave $E_{t-ref}$ (1), and, as explained in great detail in [29], is precisely what the expression for polarization would be if the relative permittivity of the dielectric medium were modulated as

$$\varepsilon_d(\omega,t) = \varepsilon_d(\omega) + \Delta\varepsilon\cos(2\omega t) \tag{5}$$

where the modulation depth is

$$\Delta\varepsilon = 2\chi_{14}^{(2)} E_{pump} \tag{6}$$

To open the large bandgap in the momentum space, this permittivity change should be commensurate with the steady-state permittivity of the effective medium, which we now proceed to evaluate and optimize.

Since our period $a = 200nm$ is much smaller than the probe wavelength, we can employ effective medium theory. For radiation polarized parallel to the layers, the effective permittivity is given by

$$\varepsilon_{probe} = \frac{d_1}{a}\varepsilon_d + \frac{d_2}{a}\varepsilon_m \tag{7}$$

where r $\varepsilon_d = 8.9$ and f $\varepsilon_m = -220 + 4i$ [31]represent the permittivities of the semiconductor and silver layers. While an ideal approach might target an effective permittivity near zero, this regime suffers from severely reduced group velocities and drastically shortened absorption lengths. To circumvent these losses, we instead target a real effective permittivity of approximately 0.7. This condition is satisfied by an $Al_{0.75}Ga_{0.25}As$ thickness of 193 nm paired with a 7 nm silver layer yielding $\varepsilon_{probe} = 0.7 + 0.14i$ .Because the metallic fraction remains minimal, the imaginary part of $\varepsilon_{probe}$ stays exceptionally low, yielding an absorption length of 2.14 $\mu$m. Consequently, roughly 40% of the incident probe light is transmitted through a 2μm sample.

Furthermore, analyzing the dispersion yields an expected large group index of $n_g = n - \lambda dn/d\lambda \sim 11.9$, corresponding to a transit time of approximately 74 fs across the 2 $\mu$m thickness. This ensures that a 50 fs probe pulse fits completely within the sample volume before the pump is activated to initiate the photonic time crystal (PTC)formation. Note that for ITO the absorption length at the same probe wavelength is only about 0.5μm which would require thinner medium and transit times which are too short to keep the probe pulse inside and observe PTC formation.

As for the pump, polarized normal to the layers, its effective permittivity can be found as

$$\frac{1}{\varepsilon_{pump}} = \frac{d_1}{a}\frac{1}{\varepsilon_d} + \frac{d_2}{a}\frac{1}{\varepsilon_m} \tag{8}$$

It yields $\varepsilon_{probe} \approx 10.301 + i0.001$ indicating negligible absorption of pump (absorption length of a mm) meaning that heating will be determined by multiphoton processes inside the semiconductor.. Unlike TCO absorption is absolutely not necessary to cause permittivity change. This ability to achieve extremely low linear absorption by separating the polarization states of the pump and probe is the salient advantage of this scheme over TCOs, which fundamentally rely on linear absorption to drive the permittivity change.

With that, I can finally estimate the depth of modulation according to (6) which for pump intensity of 1TW/cm$^2$ and $\chi_{41}^{(2)} \sim 100 pm/V$ [32] amounts to $\Delta\varepsilon_{pump} \sim 0.3$ and fractional bandgap $\Delta k / k = \Delta\varepsilon_{probe} / 2\varepsilon_{probe}^{3/2} \sim 0.26$ which should clearly be observable. Furthermore, according to (1) the parametric gain exceeds the loss, and the net temporal gain coefficient is

$$G_t = \frac{\omega}{2\varepsilon_{probe}} \left( \frac{\Delta\varepsilon_{probe}}{2} - \varepsilon'' \right) \sim 7 \times 10^{12} s^{-1} \tag{9}$$

indicating overall gain of about 50% over 70fs time the probe light stays inside the sample-something definitely not attainable in TCOs.

Even if the pump power density is reduced by an order of magnitude to 100GW/cm$^2$ would the momentum bandgap remains a very respectable 8% i.e. observable. which is well within the threshold of experimental observation. Moreover, transitioning to even shorter, single-cycle pulses should enable the observation of true time refraction and time reflection—phenomena that, up to now, have primarily been realized in the microwave domain[33].

Having come up with the scheme for potential PTC observation, despite being a theorist, I still want to rely upon hard data rather than wishful thinking and address practical obstacles on the way to implementation.

**The first issue** that arises is whether this structure can be fabricated. Here one should be humble, but still cautiously optimistic. With only 10 periods, it can in principle be fabricated using a carrier-assisted sequential chiplet transfer process: [34-36] first growing a single 193-nm $Al_{0.75}Ga_{0.25}As$ epitaxial layer on top of a sacrificial layer on a GaAs substrate, coating it with Ag, and slicing the wafer into individual chiplets. These chiplets can then be stacked vertically via layer-by-layer bonding, with the handle substrate and sacrificial layer selectively removed after each transfer step to avoid handling fragile unbacked membranes.

**The second** practical consideration is whether the effect can be observed more than once—specifically, whether the sample will evaporate after only a single pump pulse or a few, an issue often overlooked by many prominent theorists. Here, we are helped by all the design considerations built into our structure: the choice of nonlinear material assures the absence of two-photon absorption at the pump wavelength, while the chosen pump polarization ensures that the pump field hardly penetrates the strongly absorbing metal layers, which are exceptionally thin to begin with.

I estimate that a single 50-fs, 1-TW/cm$^2$ pulse will raise the temperature of the Ag layers by only about 3 K. Between pulses in a typical experiment, this heat will spread into the larger volume of the AlGaAs matrix, meaning at least 1,000 pulses would be needed to raise the overall temperature by 10 K, after which thermal diffusion reaches a steady state and halts further temperature accumulation. This robust thermal behavior stands in stark contrast to transparent conductive oxides (TCOs), where rapid local heating can easily lead to material damage.

**The third** question one needs to ask is whether the time-reflected and time-refracted signals will have enough strength to be detected. Here, we benefit from the relatively low loss of Ag at the 2300nm wavelength, which means that nearly 50% of the input probe power will exit the sample even before counting parametric amplification.

**The fourth** question is closely related to the third: in order to observe parametric time-crystals (PTC), the pump must interact with the probe for at least a few optical cycles. This is likewise assured by the low loss in the metamaterial, which allows us to scale the thickness to 2 $\mu$m and the interaction dwell time to 70fs roughly 10 optical cycles at the probe frequency. Once again, in transparent conductive oxide (TCO) experiments, the active layer thickness may be sufficient to observe some degree of time reflection, but not a true PTC.

**The fifth** issue relates to alternative means of expanding the momentum bandgap, such as the recently proposed method of exploiting strong dispersion in resonant structures[37] (the slow-light effect). Aside from the fact that the proposed structure is hardly easier to fabricate than the one introduced here, and that critical material issues like optical loss are left unaddressed—leaving permittivity modulation to be achieved seemingly by sheer willpower—one should note that while the momentum gap indeed widens, the frequency range over which the PTC effect can be observed remains severely constrained as $\Delta\omega/\omega \sim \Delta\varepsilon/2\varepsilon$. Consequently, this narrow operational bandwidth—coupled with parametric gain that fails to be enhanced relative to the crippling effects of loss—fundamentally limits the ability to actually observe PTC, even assuming that proposed contraption could be successfully fabricated.

At this point, one might be tempted to ask a **sixth question**: are all the intricate considerations that went into this proposal—which will be augmented by many others on the path to experimental demonstration—worth the trouble? Answering that definitively is beyond my level of competence. However, because the quest for PTC and a broader range of time-modulated phenomena has acquired a certain degree of importance, if not urgency, in the photonics community, I believe it is vital at least to point that community toward alternative methods of achieving it rather than remaining confined to standard TCO approaches.

So, to summarize, in this work I have suggested using a stronger, second-order nonlinearity to achieve the demonstration of PTC in the optical range. To me, $\chi^{(2)}$ is a natural choice—being the strongest (lowest order) available nonlinearity and by its very nature always ultrafast, which is why it is relied upon in all parametric devices. Why should PTC be any different? Hence, to achieve a large relative permittivity modulation, an anisotropic ENZ metamaterial incorporating alternating layers of semiconductor and silver has been proposed and its performance estimated and found superior to ENZ material based on TCOs. By separating the polarizations of the pump and probe, very low absorption of the probe can be achieved, effectively preventing thermal damage. At the same time, the relatively low optical loss for the probe permits the few-cycle interaction times necessary for PTC observation, while ensuring sufficient output power to detect all time-reflected and time-refracted signals. Whether the community warms up to this idea remains to be seen, but I can certainly hope.

## Notes

The author declare no competing financial interest.

## Data availability

The data that support the findings of this study are available from the corresponding author upon reasonable request.

## Acknowledgment

The Author acknowledges truly elucidating discussions with Dr. S. Artois of ECE department, JHU.

## References

1. E. Galiffi, R. Tirole, S. Yin, H. Li, S. Vezzoli, P. Huidobro, M. Silveirinha, R. Sapienza, A. Alù, and J. Pendry, "Photonics of time-varying media," Advanced Photonics **4**, 014002 (2022).
2. V. Pacheco-Peña, D. M. Solís, and N. Engheta, "Time-varying electromagnetic media: opinion," Opt. Mater. Express **12**, 3829-3836 (2022).
3. N. Engheta, "Four-dimensional optics using time-varying metamaterials," Science **379**, 1190-1191 (2023).
4. L. Bar-Hillel, A. Dikopoltsev, A. Kam, Y. Sharabi, O. Segal, E. Lustig, and M. Segev, "Time refraction and time reflection above critical angle for total internal reflection," Physical Review Letters **132**, 263802 (2024).
5. E. Lustig, O. Segal, S. Saha, C. Fruhling, V. M. Shalaev, A. Boltasseva, and M. Segev, "Photonic time-crystals - fundamental concepts [Invited]," Optics Express **31**, 9165-9170 (2023).
6. S. Saha, O. Segal, C. Fruhling, E. Lustig, M. Segev, A. Boltasseva, and V. M. Shalaev, "Photonic time crystals: a materials perspective [Invited]," Optics Express **31**, 8267-8273 (2023).
7. A. Boltasseva, V. Shalaev, and M. Segev, "Photonic time crystals: from fundamental insights to novel applications: opinion," Opt. Mater. Express **14**, 592-597 (2024).

8. S. Vezzoli, V. Bruno, C. DeVault, T. Roger, V. M. Shalaev, A. Boltasseva, M. Ferrera, M. Clerici, A. Dubietis, and D. Faccio, "Optical time reversal from time-dependent epsilon-near-zero media," Physical review letters **120**, 043902 (2018).
9. M. M. Asgari, P. Garg, X. Wang, M. S. Mirmoosa, C. Rockstuhl, and V. Asadchy, "Theory and applications of photonic time crystals: a tutorial," Advances in optics and photonics **16**, 958-1063 (2024).
10. Z. Hayran, J. B. Khurgin, and F. Monticone, "ℏω versus ℏk: dispersion and energy constraints on time-varying photonic materials and time crystals," Opt. Mater. Express **12**, 3904-3917 (2022).
11. H. Shim, F. Monticone, and O. D. Miller, "Fundamental Limits to the Refractive Index of Transparent Optical Materials," Adv Mater **33**(2021).
12. J. B. Khurgin, "Expanding the Photonic Palette: Exploring High Index Materials," Acs Photonics **9**, 743-751 (2022).
13. J. B. Khurgin, "Energy and power requirements for alteration of the refractive index," Laser & Photonics Reviews **18**, 2300836 (2024).
14. J. B. Khurgin and N. Kinsey, "“Nonperturbative Nonlinearities”: Perhaps Less than Meets the Eye," ACS photonics **11**, 2874-2887 (2024).
15. N. Kinsey and J. B. Khurgin, "On the bounds of index tuning in transparent conducting oxides: opinion," Opt. Mater. Express **15**, 1273-1277 (2025).
16. M. Z. Alam, I. D. Leon, and R. W. Boyd, "Large optical nonlinearity of indium tin oxide in its epsilon-near-zero region," Science **352**, 795-797 (2016).
17. O. Reshef, I. De Leon, M. Z. Alam, and R. W. Boyd, "Nonlinear optical effects in epsilon-near-zero media," Nature Reviews Materials **4**, 535-551 (2019).
18. N. Kinsey, C. DeVault, J. Kim, M. Ferrera, V. Shalaev, and A. Boltasseva, "Epsilon-near-zero Al-doped ZnO for ultrafast switching at telecom wavelengths," Optica **2**, 616-622 (2015).
19. L. Caspani, R. P. M. Kaipurath, M. Clerici, M. Ferrera, T. Roger, J. Kim, N. Kinsey, M. Pietrzyk, A. Di Falco, V. M. Shalaev, A. Boltasseva, and D. Faccio, "Enhanced Nonlinear Refractive Index in $\ensuremath{\epsilon}$-Near-Zero Materials," Physical Review Letters **116**, 233901 (2016).
20. J. B. Khurgin, M. Clerici, and N. Kinsey, "Fast and Slow Nonlinearities in Epsilon-Near-Zero Materials," Laser & Photonics Reviews **15**, 2000291 (2021).
21. N. Kinsey and J. Khurgin, "Nonlinear epsilon-near-zero materials explained: opinion," Opt. Mater. Express **9**, 2793-2796 (2019).
22. R. Secondo, J. Khurgin, and N. Kinsey, "Absorptive loss and band non-parabolicity as a physical origin of large nonlinearity in epsilon-near-zero materials," Opt. Mater. Express **10**, 1545-1560 (2020).
23. A. C. Harwood, S. Z. Lim, T. Raziman, Y. Li, J. Stones, J. W. Tisch, S. A. Horsley, S. Vezzoli, J. B. Pendry, and R. Sapienza, "Intraband and Interband Competition Drives Ultrafast Modulations of Indium Tin Oxide," arXiv preprint arXiv:2605.21432 (2026).
24. M. G. Ozlu, C. Fruhling, I. Hoffman, J. I. Choi, M. Ferrera, A. Boltasseva, and V. M. Shalaev, "Nonlinearity Reversal in Epsilon-Near-Zero Indium Tin Oxide Driven by Few-Cycle Light Pulse," arXiv preprint arXiv:2606.11567 (2026).
25. R. Maas, J. Parsons, N. Engheta, and A. Polman, "Experimental realization of an epsilon-near-zero metamaterial at visible wavelengths," Nature Photonics **7**, 907-912 (2013).
26. N. Engheta, "Metamaterials with high degrees of freedom: space, time, and more," Nanophotonics **10**, 639-642 (2021).
27. Y. J. Ding, S. J. Lee, and J. B. Khurgin, "Transversely Pumped Counterpropagating Optical Parametric Oscillation and Amplification," Physical Review Letters **75**, 429-432 (1995).
28. L. Lanco, S. Ducci, J. P. Likforman, X. Marcadet, J. A. W. van Houwelingen, H. Zbinden, G. Leo, and V. Berger, "Semiconductor Waveguide Source of Counterpropagating Twin Photons," Physical Review Letters **97**, 173901 (2006).
29. J. B. Khurgin, "Photonic Time Crystals and Parametric Amplification: Similarity and Distinction," ACS Photonics (2024).

30. V. Berger and C. Sirtori, "Nonlinear phase matching in THz semiconductor waveguides," Semiconductor science and technology **19**, 964 (2004).
31. A. D. Rakić, A. B. Djurišić, J. M. Elazar, and M. L. Majewski, "Optical properties of metallic films for vertical-cavity optoelectronic devices," Appl. Opt. **37**, 5271-5283 (1998).
32. E. Mobini, D. H. G. Espinosa, K. Vyas, and K. Dolgaleva, "AlGaAs Nonlinear Integrated Photonics," Micromachines **13**, 991 (2022).
33. H. Moussa, G. Xu, S. Yin, E. Galiffi, Y. Ra'di, and A. Alù, "Observation of temporal reflection and broadband frequency translation at photonic time interfaces," Nature Physics (2023).
34. A. Plößl and G. Kräuter, "Wafer direct bonding: tailoring adhesion between brittle materials," Materials Science and Engineering: R: Reports **25**, 1-88 (1999).
35. Z. Ren, J. Xu, X. Le, and C. Lee, "Heterogeneous Wafer Bonding Technology and Thin-Film Transfer Technology-Enabling Platform for the Next Generation Applications beyond 5G," Micromachines **12**, 946 (2021).
36. J. Xu, Y. Du, Y. Tian, and C. Wang, "Progress in wafer bonding technology towards MEMS, high-power electronics, optoelectronics, and optofluidics," International Journal of Optomechatronics **14**, 94-118 (2020).
37. X. Wang, P. Garg, M. S. Mirmoosa, A. G. Lamprianidis, C. Rockstuhl, and V. S. Asadchy, "Expanding momentum bandgaps in photonic time crystals through resonances," Nature Photonics **19**, 149-155 (2025).